\documentstyle[12pt,epsf,amsfonts,amssymb]{article}
\thicklines
\begin{document}
\begin{center}
{\bf\sc Perturbative fragmentation of vector colored particle 
into bound states with a heavy antiquark}\\
\vspace*{4mm}
V.V.Kiselev\footnote{E-mail: kiselev@mx.ihep.su}, 
A.E.Kovalsky\footnote{Moscow Institute of Physics and Technology,
Dolgoprudny, Moscow region. }\\
\vspace*{3mm}
Russian State Research Center "Institute for High Energy Physics",
Protvino,
Moscow Region,
142284,
Russia.
\end{center}
\vspace*{2mm}
\begin{abstract}
The fragmentation function of vector particle into possible bound S-wave states
with a heavy antiquark is calculated in the leading order of perturbative QCD
for the high energy processes at large transverse momenta with the different
behaviour of anomalous chromomagnetic moment. One-loop equations are derived
for
the evolution of fragmentation function moments, which is caused by the
emission of hard gluons by the vector particle. The integrated probabilities of
fragmentation are given. The distribution of bound state over the transverse
momentum with respect to the axis of fragmentation is calculated in the scaling
limit.
\end{abstract}

\section{Introduction}
An interesting problem concerning properties of interaction beyond the Standard
Model is to study production of hadrons containing leptoquarks \cite{Lept}
which are scalar and vector color-triplet particles appearing in Grand
Unification Theories, provided that their total width is much less than the
QCD-confinement scale, $\Gamma_{LQ} \ll \Lambda_{QCD}$. Recently  the
production of $(\bar{q}LQ)$--baryons in the case of scalar leptoquark was
discussed \cite{Sclep}.

In this work we study the high energy production of heavy leptoquarkonium
containing a vector leptoquark. These results can be used to calculate the
fragmentation of vector local diquarks into baryons (a similar approach was
applied to the production of $\Omega_{ccc}$ in \cite{Omega}).
For the sake of convenience the local color-triplet vector field will be
referred to as the leptoquark in this paper.

New problem arising in this case is a choice of the lagrangian for the  vector
leptoquark interaction with gluons. Indeed, to the lagrangian of a free vector
field 
$$-1/2 H_{\mu \nu} \bar{H}^{\mu \nu},
$$ 
where $H_{\mu \nu}= \partial_{\mu} U_{\nu}-\partial_{\nu} U_{\mu}$,  
$U_{\mu}$ is the vector complex field with derivatives substituted by covariant
ones, we can add the gauge invariant term proportional to
$$
S^{\alpha \beta}_{\mu \nu} G^{\mu \nu} U_{\beta} \bar{U}_{\alpha},
$$
where $S^{\alpha \beta}_{\mu \nu}=1/2(\delta^{\alpha}_{\mu}
\delta^{\beta}_{\nu}-\delta^{\alpha}_{\nu} \delta^{\beta}_{\mu})$
is the tensor of spin, $G^{\mu \nu}$ is the gluon field strength tensor.
It leads to the appearance of a parameter in the gluon--leptoquark
vertex (the so-called anomalous chromomagnetic moment, see Section 2).
In this work we discuss the production of a $1/2$-spin bound state 
containing the heavy vector particle at various values of this parameter.

At high transverse momenta, the dominant production mechanism for the heavy
leptoquarkonium bound states is
the leptoquark fragmentation,
which can be calculated in perturbative QCD \cite{Bra-rev} after the isolation
of soft binding factor extracted from the non-relativistic potential models
\cite{Martin, Buchm}.
The corresponding fragmentation function is universal for any high energy
process for the direct production of leptoquarkonia.

In the leading $\alpha_s$-order,
the fragmentation function has a scaling form,
which is the initial condition for the perturbative QCD evolution caused by the
emission of hard gluons by the leptoquark before the hadronization.
The corresponding splitting function differs from that for the heavy quark
because of the spin structure of gluon coupling to the leptoquark.

In this work the scaling fragmentation function is calculated in Section 2 for
two different cases of the anomalous chromomagnetic moment behaviour. The limit
of infinitely heavy leptoquark, $m_{LQ}\to \infty$, is obtained from the full
QCD consideration for the fragmentation. The distribution of bound state over
the transverse momentum with respect to the axis of fragmentation is calculated
in the scaling limit in Section 3. The splitting kernel of the DGALAP-evolution
is derived in Section 4, where the one-loop equations of renormalization group
for the moments of fragmentation function are obtained and solved. These
equations are universal, since they do not depend on whether the leptoquark
will bound or free at low virtualities, where the perturbative evolution stops.
The integrated probabilities of leptoquark fragmentation into the heavy
leptoquarkonia are evaluated in Section 5. The results are summarized in
Conclusion.
 
 \section{Fragmentation function in leading order} 

The contribution of fragmentation into direct production of heavy
leptoquarkonium has the form
$$
d\sigma[l_H(p)] = \int_0^1 dz\; d\hat \sigma[LQ(p/z), \mu]\; 
D_{LQ\to l_H}(z, \mu), 
$$
where $d\sigma$ is the differential cross-section for the production of
leptoquarkonium with the 4-momen\-tum $p$, $d\hat \sigma$  is that of the hard
production of leptoquark with the scaled momentum $p/z$, and D is interpreted
as the fragmentation function depending on the fraction of momentum carried out
by the bound state. The value of $\mu$ determines the factorization scale. In
accordance with the general DGLAP-evolution, the $\mu$-dependent fragmentation
function satisfies the equation
\begin{equation}
\frac{\partial D_{LQ\to l_H}(z, \mu)}{\partial \ln \mu} =
\int_z^1 \frac{dy}{y} \; P_{LQ\to LQ}(z/y, \mu)\; D_{LQ\to l_H}(y, \mu), 
\label{DGLAP}
\end{equation}
where P is the kernel caused by emission of hard gluons by the leptoquark
before the production of heavy quark pair. Therefore, the initial form of
fragmentation function is determined by the diagram shown in Fig.\ref{diag},
and, hence, the corresponding initial factorization scale is equal to $\mu=
2m_Q$. Furthermore, this function can be calculated as an expansion in
$\alpha_s(2m_Q)$. The leading order contribution is evaluated in this Section.

\setlength{\unitlength}{1mm}
\begin{figure}[th]
\begin{center}
\begin{picture}(100,80)
\put(0,0){\epsfxsize=9cm \epsfbox{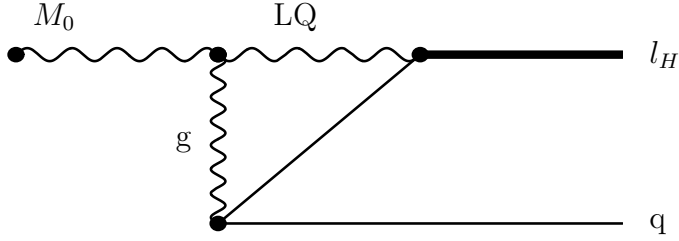}}
\put(90,8){q}
\put(90,30){$l_H$}
\put(40,35){LQ}
\put(27,19){g}
\put(8,35){$M_0$}
\end{picture}
\end{center}
\caption{The diagram of leptoquark fragmentation into the heavy
leptoquarkonium.}
\label{diag}
\end{figure}

Consider the fragmentation diagram in the system, where the momentum of initial
leptoquark has the form $q=(q_0, 0, 0, q_3)$ and the leptoquarkonium one is
$p$, so that 
$$
q^2=s,  \;\; p^2=M^2.\;\; 
$$
In the static approximation for the bound state of leptoquark and heavy quark,
the quark mass is expressed as $m_Q= r M$, and the leptoquark mass equals
$m=(1-r)M=\bar r M$. The gluon--vector leptoquark vertex has the form
 \begin{equation}
T_{\alpha\mu\nu}^{VVg}=-i g_{s} t^{a} [g_{\mu\nu} (q+\bar r
p)_{\alpha}-g_{\mu\alpha} ((1+\ae )\bar r p-\ae  q)_{\nu}-g_{\nu\alpha}
((1+\ae )q -\ae  \bar r p)_{\mu}],
\label{VgV}
\end{equation}
where $\ae$ is the anamalous chromomagnetic moment, $t^a$ is the QCD group
generator. The sum over the vector leptoquark polarizations with the momentum
$q\; (q^2=s)$ depends on the choice of the gauge of free field lagrangian (for
example, the Stueckelberg gauge), but the fragmentation function is a physical
quantity and has not to depend on the gauge parameter changing the contribution
of longitudinal components of the vector field.
So,
the sum over polarization can be taken in the form
$$
P(q)_{\mu\nu}= -g_{\mu\nu}+\frac{q_\mu q_\nu}{s}.
$$ 
The matrix element of the fragmentation into the baryon with the spin of $1/2$
has the form  
\begin{equation}
{\cal M} = -\frac{2\sqrt{2\pi}\alpha_s}{9\sqrt{M^3}}
\frac{R(0)}{r\bar r (s-m^2)^2}
P(q)_{\nu\delta}P(\bar r p)_{\mu\eta}T_{\alpha\mu\nu}^{VVg}
\rho_{\alpha\beta}\;
\bar q \gamma^\beta(\hat p-M) \gamma^\eta \gamma^5 l_H \; {\cal M}_0^\delta, 
\label{M}
\end{equation}
where the sum over the gluon polarization is written down in the axial gauge
with
$n=(1, 0, 0, -1)$
$$
\rho_{\mu\nu}(k) = -g_{\mu\nu}+\frac{k_\mu n_\nu+k_\nu n_\mu}{k\cdot n}, 
$$
with $k=q-(1-r)p$.
The spinors $l_H$ and $\bar q$ correspond to the leptoquarkonium and heavy
quark associated to the fragmentation. ${\cal M}_0$ denotes the matrix element
for the hard production of leptoquark at high energy, $R(0)$ is the radial
wave-function at the origin. The matrix element squared and summed over the
helicities of particles in the final state has the form
$$
 \bar{ |M|}^2=W_{\mu\nu} M_0^\mu M_0^\nu
$$
In the limit of high energies $q\cdot n \to \infty$, $ W_{\mu\nu}$ behaves like
\begin{equation}
W_{\mu\nu}=-g_{\mu\nu}W+R_{\mu\nu}, 
\label{W}
\end{equation}
where $R_{\mu\nu}$ can depend on the gauge parameters and leads to scalar
formfactor terms which are small in comparison with $W$ in the limit of
$q\cdot n \to \infty$. 
Define
$$
z=\frac{p\cdot n}{q\cdot n}. 
$$
The fragmentation function is determined by the expression \cite{Bra-frag}
$$ 
D(z) = \frac{1}{16\pi^2}\int ds \theta\biggl(s-\frac{M^2}{z}-\frac{m_Q^2}
{1-z}\biggr)\;W 
, 
$$ 
where $W$ is defined in (\ref{W}). The integral in the expression for the
fragmentation function diverges logarithmically at a constant value of
anamalous chromomagnetic moment if $\ae$ does not equal $-1$. In this work we
consider two sets for the behaviour of anamalous chromomagnetic moment.

Set I is defined by $\ae=-1$. Here we observe that the obtained fragmentation
function coincides with that for the scalar leptoquark \footnote{In paper
\cite{Sclep} an arithmetic error was made, which slightly affects upon the
final result at small $r$.} except the factor of $1/3$
\begin{eqnarray}
D(z) = \frac{8\alpha_s^2}{243\pi}\;
\frac{|R(0)|^2}{M^3 r^2\bar r^2}\;
\frac{z^2(1-z)^2}{[1-\bar r z]^6} &\cdot&
\bigg\{3+3r^2-(6-10r+2r^2+2r^3)z+\nonumber \\
&&+(3-10r+14r^2-10r^3+3r^4)z^2)\bigg\}, 
\label{frscal}
\end{eqnarray}
which tends to
\begin{equation}
\tilde D(y)= \frac{8\alpha_s^2}{243\pi\ y^6}\;
\frac{|R(0)|^2}{m_Q^3}\; \frac{(y-1)^2}{r}\bigg\{8+4y+3y^2\bigg\}, 
\label{tilde}
\end{equation}
at $r\to 0$ and $y=(1-(1-r)z)/(rz)$.
The limit of $\tilde D(y)$ is in agreement with the general consideration of
$1/m$-expansion for the fragmentation function \cite{JR},
where 
$$
\tilde D(y) = \frac{1}{r}a(y) +b(y). 
$$
The dependence on $y$ in $a(y)$ is the same as for the fragmentation of the
heavy quark into the quarkonium \cite{Bra-frag}.

Set II:  $\ae$ behaves like $-1+A M^2
/(s-m_{LQ}^2)$.
The obtained fragmentation function is
\begin{eqnarray}
 D(z) &=&\frac{8\alpha_s^2}{243\pi}\;
\frac{|R(0)|^2}{16 M^3 r^2\bar r^2}\;
\frac{z^2(1-z)^2}{[1-\bar r z]^6}\cdot
\nonumber \\ &&
\bigg\{16(3+3r^2-6z+10rz-2r^2z-2r^3z+
\nonumber \\ &&
+3z^2-10z^2r+14z^2r^2-10z^2r^3+3z^2r^4)+
\nonumber \\ &&
+A(3A+24r-6zA-2rzA-32rz-16r^2z+3z^2A+
\nonumber \\ &&
+2z^2rA+8z^2r+3z^2r^2A-32z^2r^2+24z^2r^3)
\bigg\}.
\label{frvect}
\end{eqnarray}
It tends to
\begin{eqnarray}
 \tilde D(y) &=&\frac{8\alpha_s^2}{243 \pi\ 16 y^6}\;
\frac{|R(0)|^2}{m_Q^3}\; \frac{(y-1)^2}{r}\
\nonumber\\ &&
\bigg\{16(8+4y+3y^2)+A(8A-8yA+3y^2A-64+16y) \bigg\},
\end{eqnarray}
at $r\to 0$ and $y=(1-(1-r)z)/(rz)$.
The perturbative functions in the leading $\alpha_s$-order are shown in
Fig.\ref{d-fig} at $r=0.02$.
We see hard distributions,
which become softer with the evolution (see ref.\cite{Sclep}).

\setlength{\unitlength}{1mm}
\begin{figure}[t]
\begin{center}
\begin{picture}(100, 80)
\put(0, 0){\epsfxsize=11cm \epsfbox{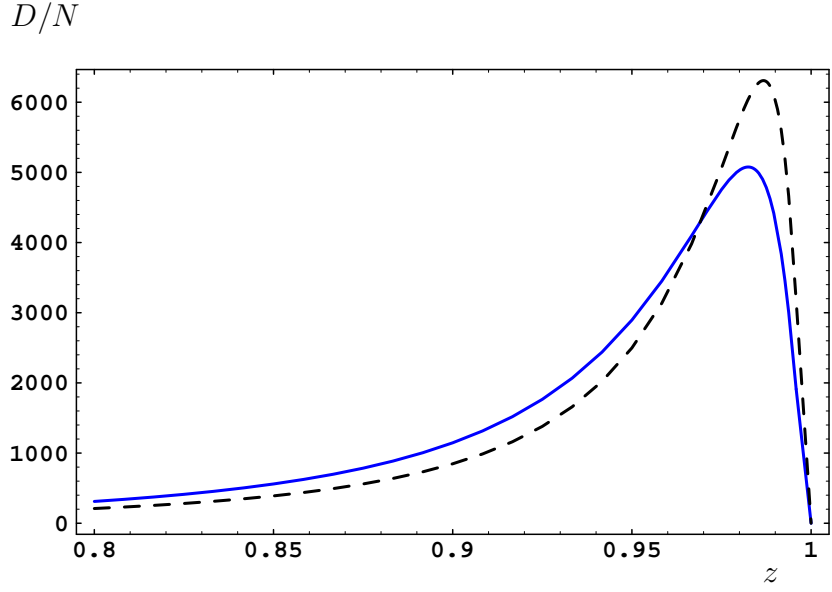}}
\put(100, 6){$z$}
\put(0, 80){$D/N$}
\end{picture}
\end{center}
\caption{The fragmentation function of leptoquark into the heavy
leptoquarkonium, the $N$-factor is determined by
$N=\frac{8\alpha_s^2}{243\pi}\; \frac{|R(0)|^2}{M^3 r^2(1-r)^2}$,  the
fragmentation function for Set I is shown by the dashed line, the fragmentation
function for Set II ($A=3$) is given by the solid line at $r=0.02$.} 
\label{d-fig}
\end{figure}

\section{Transverse momentum of leptoquarkonium}

 In the system with an infinite momentum of the fragmentating leptoquark its
invariant mass is expressed by the fraction of longitudinal leptoquark
momentum $z$ and transverse momentum with respect to the fragmentation axis
$p_T$ (see Fig. \ref{diag}) as 
$$
s = m^2 + \frac{M^2}{z(1-z)}[(1-(1-r)z)^2+t^2], 
$$
where $t=p_T/M$. The calculation of diagram in Fig. \ref{diag} gives the double
distribution
$$
\frac{d^2 P}{ds\; dz} = {\cal D}(z, s), 
$$
where the Set I function ${\cal D}$ has the form 
\begin{eqnarray}
{\cal D}(z, s) &=& \frac{256\alpha_s^2}{81 \pi} \;
\frac{|R(0)|^2}{r^2\bar r^2}\;
\frac{M^3}{[1-\bar r z]^2 (s-m^2)^4} \nonumber\\ &&
\biggl\{r\bar r^2+\bar r (1+r-z(1+4r-r^2))\frac{s-m^2}{M^2}-
 z(1-z) \biggl(\frac{s-m^2}{M^2}\biggr)^2\biggr\}. 
\end{eqnarray}
For Set II,
we find
\begin{eqnarray}
{\cal D}(z, s) &=& \frac{8\alpha_s^2}{81 \pi} \; \frac{|R(0)|^2}{r^2\bar r^2}\;
\frac{M^3}{[1-\bar r z]^2 (s-m^2)^4}
\nonumber\\ &&
\biggl\{8r(A-4+4r)^2(1-z+rz)^2+
\nonumber \\ &&
 +2(-A-4-4r+zA+4z-rzA+16rz-4r^2z)
 \nonumber \\ &&
 (1-z+rz)(A-4+4r)\frac{s-m^2}{M^2} 
 -32(1-z)z \biggl(\frac{s-m^2}{M^2}\biggr)^2 \biggr\}. 
\end{eqnarray}
The distribution of leptoquarkonium over the transverse momentum can be
obtained by the integration over $z$
 $$
D(t) = \int_0^1 dz\; {\cal D}(z, s)\; \frac{2M^2 t}{z(1-z)}. 
$$
 For Set I we have 
\begin{eqnarray}
D(t) &=& \frac{64\alpha_s^2}{81 \pi} \; \frac{|R(0)|^2}{3(1-r)^5 M^3}\;
\frac{1}{t^6}\;
\nonumber \\ &&
\bigg\{t (30r^3-30r^4-61t^2r+45r^2t^2+33r^3t^2-
\nonumber \\ &&
-17r^4t^2+3t^4-9rt^4+15 r^2t^4-9r^3t^4)-
\nonumber\\ &&
(30r^4-99r^2t^2-54r^3t^2+27r^4t^2+9t^4+18rt^4-6r^2t^4+
\nonumber \\ &&
+18r^3t^4+3r^4t^4+3t^6-6rt^6+9r^2t^6){\rm arctg}
\biggl(\frac{(1-r)t}{r+t^2}\biggr)+
\nonumber \\ &&
24(2r^3t+rt^3+r^2t^3)\;\ln\biggl(\frac{r^2(1+t^2)}{r^2+t^2}\biggr)\bigg\}. 
\end{eqnarray}
The distribution for Set II is given in Appendix.
The typical form of distribution over the transverse momentum is shown in Fig.
\ref{dt-fig}.

\setlength{\unitlength}{1mm}

\begin{figure}[t]
\begin{center}
\begin{picture}(110, 70) 
\put(0, 0){\epsfxsize=14cm \epsfbox{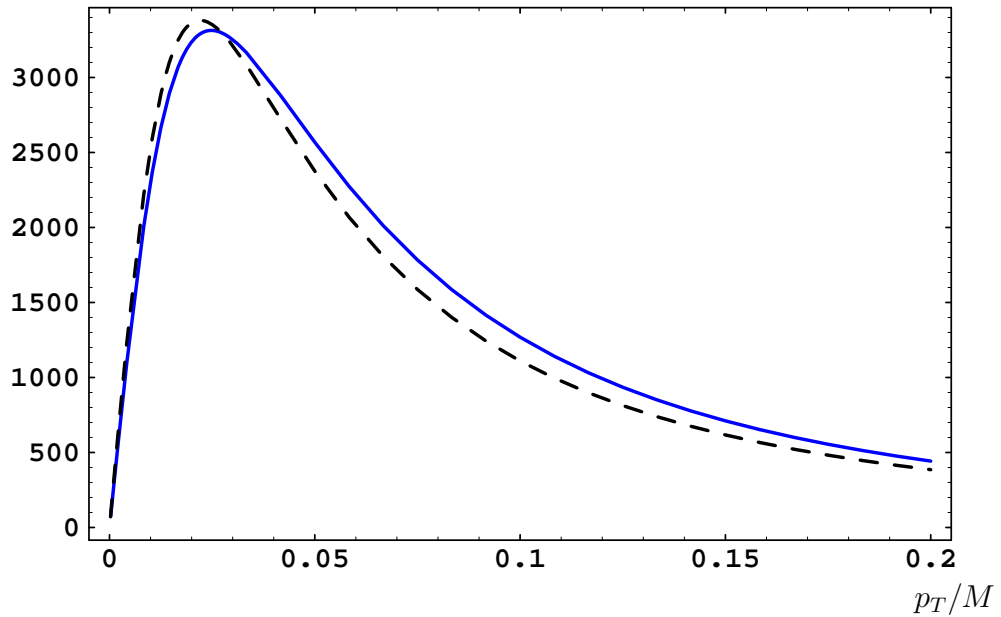}}
\put (120, 0){$p_T/M$}
\put (0, 90){$D(p_T)/N_t$}
\end{picture}
\end{center}
\caption{The distributions over the transverse momentum with respect to the
axis of fragmentation,
$N_t$-factor is determined by 
$N_t=\frac{8\alpha_s^2}{81\pi}\; \frac{|R(0)|^2}{M^4 r^2(1-r)^7}$.
The dashed line represents the result for Set I, the solid line does it for Set
II.}
\label{dt-fig}
\end{figure}

\section{Hard gluon emission}
The one-loop contribution can be calculated in the way described in the
previous Sections. This term does not depend on the part of leptoquark-gluon
vertex with the anomalous chromomagnetic moment, therefore the splitting kernel
coincides with that for the scalar leptoquark. It equals
\begin{equation}
P_{LQ\to LQ}(x,\mu) = \frac{4\alpha_s(\mu)}{3\pi}\;
\bigg[\frac{2x}{1-x}\biggr]_+,
\label{P}
\end{equation}
where the "plus" denotes the ordinary action: $\int_0^1 dx f_+(x)\cdot g(x)=
\int_0^1 dx f(x)\cdot [g(x)-g(1)]$. The scalar leptoquark splitting function
can be compared with that of the heavy
quark 
$$
P_{Q\to Q}(x,\mu) = \frac{4\alpha_s(\mu)}{3\pi}\;
\bigg[\frac{1+x^2}{1-x}\biggr]_+,
$$
which has the same normalization factor at $x\to 1$. Furthermore, multiplying
the evolution equation by $z^n$ and integrating over $z$, one can get from
(\ref{DGLAP}) the $\mu$-dependence of moments $a_{(n)}$ up to the one-loop
accuracy of renormalization group
 \begin{equation}
\frac{\partial a_{(n)}}{\partial \ln \mu} = - \frac{8\alpha_s(\mu)}{3\pi}\;
\bigg[\frac{1}{2}+\ldots +\frac{1}{n+1}\biggr]\; a_{(n)}, \;\;\; n\ge 1.
\label{da}
\end{equation}
At $n=0$ the right hand side of (\ref{da}) equals zero, which means that the
integrated probability of leptoquark fragmentation into the heavy
leptoquarkonium is not changed during the evolution, and it is determined by
the initial fragmentation function calculated perturbatively in Section 2.

The solution of equation (\ref{da}) has the form 
\begin{equation}
a_{(n)}(\mu) = a_{(n)}(\mu_0)\; \biggl[\frac{\alpha_s(\mu)}
{\alpha_s(\mu_0)}\biggr]^{\frac{16}{3\beta_0}
\bigg[\frac{1}{2}+\ldots +\frac{1}{n+1}\biggr]},
\label{a}
\end{equation}
where one has used the one-loop expression for the QCD coupling constant
$$
\alpha_s(\mu) = \frac{2\pi}{\beta_0\ln(\mu/\Lambda_{QCD})},
$$
where $\beta_0= 11-2 n_f/3$, with $n_f$ being the number of quark flavors with
$m_q<\mu<m_{LQ}$. Relation (\ref{a}) is universal one, since it is independent
of whether the leptoquark is free or bound at the virtualities less than
$\mu_0$. We can use the evolution for the fragmentation into the heavy
leptoquarkonium. The leptoquark can lose about 20\% of its momentum before the
hadronization \cite{Sclep}.

\section{Integrated probabilities of fragmentation}

\begin{figure}[t]
\hspace*{3cm}
\begin{center}
\begin{picture}(110, 130)
\put(0, 0){\epsfxsize=13cm \epsfbox{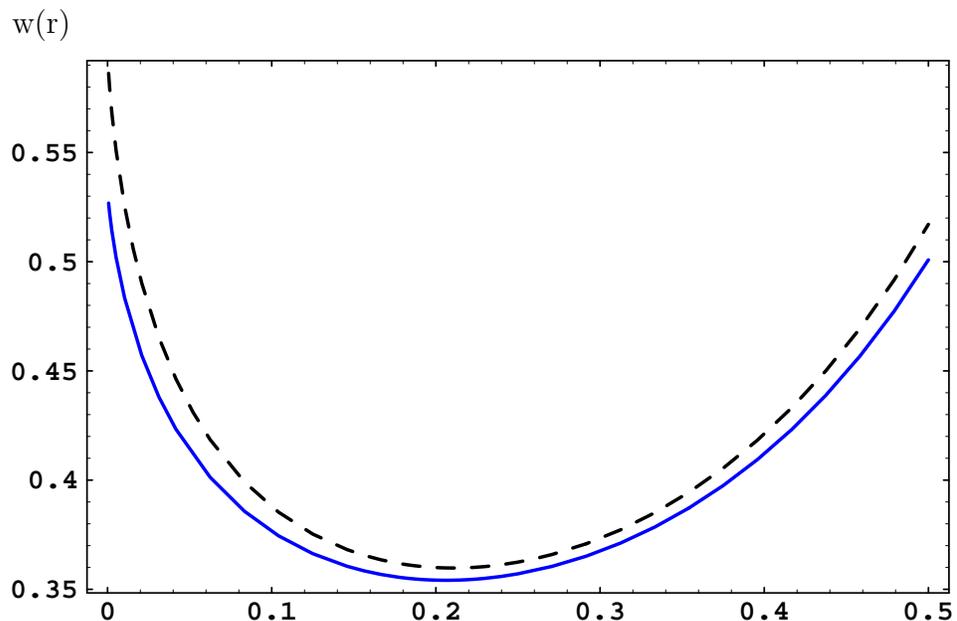}}
\put (130, 5){r}
\put (0, 90){$\rm{w(r})$}
\end{picture}
\end{center}
\caption{The w-functions for the leptoquark fragmentation into the heavy
leptoquarkonium versus the fraction $r=m_Q/M$. The curves correspond to Sets as
in Fig. \ref{dt-fig}.}
\label{w-fig}
\end{figure}

As has been mentioned above, the evolution conserves the integrated probability
of fragmentation which can be calculated explicitly
\begin{eqnarray}
\int dz\; D(z) &=& \frac{8\alpha_s^2}{81\pi}\; \frac{|R(0)|^2}{16 m_Q^3}\; 
{\rm w}(r).
\end{eqnarray}
For Set I, we have
 \begin{equation}
{\rm w}(r)=  \frac{16[(8+15r-60r^2+100r^3-60r^4-3r^5)+
30r(1-r+r^2+r^3)\ln r]}{15(1-r)^7}. 
\end{equation}
 For Set II, we find $(A=3)$ 
\begin{eqnarray}
{\rm w}(r) &=& 
\bigg\{(143+701r-1882r^2+3250r^3-3245r^4+2017r^5-936r^6-48r^7)+
\nonumber \\ &&
+30r(25-21r+43r^2+r^3+8r^4+16r^5)\ln r\bigg\}\frac{1}{15(1-r)^9}.
\end{eqnarray}
The ${\rm w}(r)$ functions are shown in Fig. \ref{w-fig} at low $r$.

\section{Conclusion}
 
In this work the dominant mechanism for the production of bound states of spin
$1/2$ of a local color-triplet vector field with a heavy antiquark is
considered for high energy processes at large transverse momenta, where the
fragmentation contributes as the leading term. We investigate two sets of the
anamalous chromomagnetic moment behaviour. Set I is defined by $\ae =-1$ (the
expression for the fragmentation function diverges logarithmically at a
constant value of anamalous chromomagnetic moment if $\ae$ is not equal to
$-1$). Here we observe that the obtained fragmentation function coincides with
that for the scalar leptoquark up to the factor of $1/3$
\begin{eqnarray}
D(z) = \frac{8\alpha_s^2}{243\pi}\;
\frac{|R(0)|^2}{M^3 r^2\bar r^2}\;
\frac{z^2(1-z)^2}{[1-\bar r z]^6} &\cdot&
\bigg\{3+3r^2-(6-10r+2r^2+2r^3)z+\nonumber \\
&&+(3-10r+14r^2-10r^3+3r^4)z^2)\bigg\}, 
\end{eqnarray}
where $r$ is the ratio of heavy quark mass to the mass of the bound state. In
the infinitely heavy leptoquark limit, $D(z)$ has the form, which agrees with
what expected from general consideration of $1/m$-expansion for the
fragmentation function. Set II is defined by $\ae=-1+A M^2/(s-m^2_{LQ})$. The
fragmentation function for Set II differs from that for Set I: 
\begin{eqnarray}
 D(z) &=&\frac{8\alpha_s^2}{243\pi}\;
\frac{|R(0)|^2}{16 M^3 r^2\bar r^2}\;
\frac{z^2(1-z)^2}{[1-\bar r z]^6}\cdot
\nonumber \\ &&
\bigg\{16(3+3r^2-6z+10rz-2r^2z-2r^3z+
\nonumber \\ &&
+3z^2-10z^2r+14z^2r^2-10z^2r^3+3z^2r^4)+
\nonumber \\ &&
+A(3A+24r-6zA-2rzA-32rz-16r^2z+3z^2A+
\nonumber \\ &&
+2z^2rA+8z^2r+3z^2r^2A-32z^2r^2+24z^2r^3)
\bigg\}.
\end{eqnarray}
The distribution of bound state over the transverse momentum with respect to
the axis of fragmentation is calculated in the scaling limit. The corresponding
distribution functions are given by the expression $(11)$ for Set I and the
expression from Appendix. The hard gluon corrections caused by the 
splitting of vector leptoquark are taken into account so that the evolution
kernel has the form  
 \begin{equation}
 P_{LQ\to LQ}(x,\mu) = \frac{4\alpha_s(\mu)}{3\pi}\;
 \bigg[\frac{2x}{1-x}\biggr]_+,
 \end{equation}
which results in the corresponding one-loop equation for the moments of
fragmentation functions (see eqs.(13), (14)). The numerical estimates show that
the probabilities of fragmentation into bound states with c- and b-quarks of
heavy vector leptoquark with the mass about 400 GeV are of the order of
$10^{-(3-4)}$. This suppression makes the experimental observation of such
states rather difficult. However we can use the perturbative expressions for
the model of fragmentations into doubly heavy baryons, where the integrated
probabilities are of the order of $10^{-(1-2)}$. The results will be used to
investigate the fragmentation of doubly heavy vector diquarks into baryons
elsewhere.

This work was in part supported by the Russian Foundation for Basic Research,
grants 99-02-16558 and 96-15-96575.
 
 \section{Appendix}
The expression for the distribution over the transverse
momentum for Set II ($\ae=-1+\frac{3 M^2}{s-m^2_{LQ}}$) is given by  
 \begin{eqnarray}
D(t) &=&\frac{8\alpha_s^2}{81 \pi} \; \frac{|R(0)|^2}{3(1-r)^7 M^3}\;
\frac{1}{t^6}\;
\nonumber \\ &&
\bigg\{30r^3t-270r^4t+720r^5t-480r^6t-727t^3r+2441r^2t^3-
\nonumber \\ &&
-2041r^3t^3-65r^4t^3+664r^5t^3-272r^6t^3+129t^5-285t^5r+
\nonumber \\ &&
+477r^2t^5-489r^3t^5+312r^4t^5-144r^5t^5-
(30r^4-240r^5+480r^6+\nonumber \\ &&
+t^2(-909r^2+2412r^3-45r^4-1080r^5+432r^6)+
\nonumber \\ &&
+t^4(171+288r-492r^2+696r^3-165r^4+264r^5+48r^6)+
\nonumber \\ &&
+t^6(129-156r+321r^2-168r^3+144r^4))
{\rm arctg}\biggl(\frac{(1-r)t}{r+t^2}\biggr)+
\nonumber \\ &&
+192tr(1-r)(-2t^2(1-r^2)+(1-4r)r^2)
{\rm
ln}\biggl(\frac{r^2(1+t^2)}{r^2+t^2}\biggr)\bigg\}.
\end{eqnarray}


\begin{thebibliography}{**}
\bibitem{Lept}
W.Buchm\" uller,  R.R\" uckl,  D.Wyler,  Phys. Lett. B191,
442 (1987).
\bibitem{Sclep}
V.V.Kiselev, Phys. Rev. D58, 054008 (1998);\\
V.V.Kiselev, Phys. Atom. Nucl. 62, 300 (1999) [Yad. Fiz. 62, 335 (1999)]. 
\bibitem{Omega}
V.A.Saleev, hep-ph/9906515 (1999).
\bibitem{Bra-rev}
E.Braaten,  S.Fleming,  T.C.Yuan,  Ann.  Rev.  Nucl.  Part.  Sci. 46, 197
(1997).
\bibitem{Martin}
A.Martin,  Phys.  Lett.  93B, 338 (1980).
\bibitem{Buchm}
W.Buchm\" uller,  C.H.H.Tye,  Phys.  Rev.  D24, 132 (1981).
\bibitem{Bra-frag}
E.Braaten,  T.C.Yuan,  Phys.  Rev.  Lett.  71, 1673 (1993);\\
E.Braaten,  K.Cheung,  T.C.Yuan,  Phys.  Rev. D48, 4230 (1993);\\
E.Braaten,  K.Cheung,  T.C.Yuan,  Phys.  Rev. D48, 5049 (1993).
\bibitem{JR}
R.L.Jaffe,  L.Randall,  Nucl. Phys.  B415, 79 (1994).
\end{thebibliography}
\end{document}